\documentclass[12pt,preprint]{aastex}

\def\be{\begin{equation}}
\def\ee{\end{equation}}

\def\ergs{{\rm\,erg\,s^{-1}}}
\newcommand{\msun}{{M}_{\sun}}

\catcode`\@=11 
\def\@versim#1#2{\vcenter{\offinterlineskip
        \ialign{$\m@th#1\hfil##\hfil$\crcr#2\crcr\sim\crcr } }}

\shorttitle{****} \shortauthors{****}
\begin{document}

\title{Observational evidence for AGNs feedback at parsec scale}

\author{Feng Yuan and Miao Li}
\affil{Key Laboratory for Research in Galaxies and Cosmology,
Shanghai Astronomical Observatory, Chinese Academy of Sciences, 80
Nandan Road, Shanghai 200030, China} \email{fyuan@shao.ac.cn}

\begin{abstract}
In a hot accretion flow, the radiation from the innermost region of
the flow propagates outward and heats the electrons at large radii
via Compton scattering. It has been shown in previous works that if
the radiation is strong enough, $L\ga 2\%L_{\rm Edd}$, the electrons
at the Bondi radius ($r_B\sim 10^5 r_s$) will be heated to be above
the virial temperature thus the accretion will be stopped. The
accretion will recover after the gas cools down. This results in the
oscillation of the black hole activity. In this paper we show that
this mechanism is the origin of the intermittent activity of some
compact young radio sources. Such intermittency is required to
explain the population of these sources. We calculate the timescales
of the black hole oscillation and find that the durations of active
and inactive phases are $3\times 10^4 (0.1/\alpha)(M/10^8\msun)
(L/2\%L_{\rm Edd})^{-1/2}~{\rm yr}$ and
$10^5(\alpha/0.1)(M/10^8\msun)~{\rm yr}$, respectively, consistent
with those required to explain observations. Such kind of feedback
occurring at parsec scale should be common in low-luminosity AGNs
and should be considered when we consider their matter and energy
output.

\end{abstract}


\section{Introduction: intermittent activity of compact radio sources}

Feedback from active galactic nuclei (AGNs) is now widely believed
to play an important role in the formation and evolution of
galaxies, and to be responsible for many observational results
including the $M_{\rm BH}-\sigma$ relation and the suppression of
star formation in elliptical galaxies (e.g., Silk \& Rees 1998;
Fabian 1999; King 2003; Sanonov et al. 2005; Murray et al. 2005; Di
Matteo et al. 2005; Croton et al. 2006; Ciotti \& Ostriker 2001,
2007). While significant progresses have been made, many details of
the feedback are still uncertain (Ostriker et al. 2010). One
approach to improve this situation is to look for more direct
observational evidence for feedback and investigate them carefully.

{Feedback often causes intermittent activities of AGNs}, thus it
should be helpful to look for observational evidence for such an
intermittency (see review by Czerny et al. 2009). One evidence is
the central dominant galaxies in galaxy clusters observed by {\em
Chandra X-ray Observatory} and {\em XMM/Newton}. X-ray images show
cavities and ripples that gives the evidence for repetitive
outbursts. Another example, which we will focus on in this paper, is
radio galaxies. Many case studies of radio structure directly show
evidence for two or more of active periods. Some sources are very
compact, only a few kpc in size, which indicates that they are very
young. These small radio sources are found to constitute 10\%-30\%
of all sources in a flux-limited sample. O'Dea \& Baum (1997; see
also Snellen et al. 2000) have studied the statistical properties of
a combined sample of gigahertz-peaked spectrum sources (GPS) and
compact steep spectrum radio sources (CSS). Their results show the
existence of far too many compact (young) sources in comparison with
the number of galaxies with extended old radio structures. If the
total activity period lasts for $10^8$ yr, the number of sources
with the ages below $10^3$ yr should be roughly $10^5$ times lower
than the number of sources older than $10^3$ year, which is however
not the case. The most likely explanation for the overabundance is
that the activity is intermittent, i.e., the sources undergo an
active or outburst phase lasting for $10^4$ yr which recurs every
$10^5$ yr (Reynolds \& Begelman 1997; Kaiser et al. 2000;
Kunert-Bajraszewska et al. 2005). Detection of several candidates
for dying compact sources supports this view (Giroletti et al. 2005;
Parma et al. 2007).

To explain the origin of such intermittent activity, Czerny et al.
(2009) have compiled a sample consisting of 72 GPS with measured
age. The age is determined based on their measured hot-spot
separation speed (kinematic age) or on the synchrotron cooling time.
The majority of the age values are within the range of 200 yr and
$10^4$ yr. The monochromatic luminosity $L_{\rm 5GHz}$ ($\equiv \nu
L_{\nu}$ at 5GHz) are all available for these sources, with typical
values of $10^{42}$-$10^{45} \ergs$ and average value of
$10^{43.6}\ergs$. This value can be used to estimate their
bolometric luminosities of the accretion flows, $L_{\rm acc}$, in
the following way (Czerny et al. 2009). For some sources, the 2-10
keV luminosity $L_{\rm x}$ are available. It is found that $L_{\rm
x}$ is comparable to or a factor of $\sim 5$ higher than, $L_{\rm
5GHz}$ (Vink et al. 2006; Siemiginowska et al. 2008)\footnote{The
caveat is that the observed X-ray emission may not be dominated by
the accretion flow (Stawarz et al. 2008).}. To estimate $L_{\rm
acc}$ from $L_{\rm x}$ we have to assume some template of the
spectral energy distribution (SED) of the sources, either luminous
AGNs or low-luminosity AGNs (LLAGNs), because their SEDs are
characteristically different (Ho 1999). Assuming the SEDs of these
GPS sources are similar to those of luminous AGNs, $L_{\rm acc}$ is
at least 10 times higher than $L_{\rm x}$ (Elvis et al. 1994).
Assuming a fiducial number of 50, we have $L_{\rm acc}\sim 50 L_{\rm
x}\sim 100L_{\rm 5GHz}\sim 10^{44}-10^{47}\ergs$ with averaged value
of $3\times 10^{45}\ergs$. For several sources whose black hole
masses are measured, $L_{\rm acc}\sim (0.03-1.5)L_{\rm Edd}$ with
typical value of $\sim 0.1L_{\rm Edd}$ (Czerny et al. 2009).
However, as we will argue below, these sources are more likely
low-luminosity AGNs whose SEDs are distinctively different from
luminous AGNs, characterized mainly by the lack of big-blue-bumps
(Ho 1999; 2008). In this case, $L_{\rm acc}\sim 10 L_{\rm x}$ (Ho
1999), so we have $L_{\rm acc}\sim (10-50) L_{\rm 5GHz}\sim
(10^{43}-5\times 10^{46})\ergs$. The typical Eddington ratio is then
$L_{\rm acc}\sim 0.02L_{\rm Edd}$. Another statistical result from
the sample in Czerny et al. (2009) is that the more luminous the
sources are, the younger they are.

The only model proposed so far to explain the intermittent activity
of these compact radio sources is the thermal instability of a
radiation-pressure-dominated standard thin disk (Czerny et al. 2009;
see also Janiuk \& Czerny 2011). In \S2, we discuss several problems
of this interpretation. We then propose in \S3 an alternative model,
i.e., a ``Global Compton scattering feedback" model, based on Yuan,
Xie \& Ostriker (2009). We assume that these radio sources are
powered by hot accretion flows (advection-dominated accretion flow;
ADAF) rather than the standard thin disk. The Compton heating to the
electrons at $10^5 r_s$ ($r_s\equiv 2GM/c^2$ is the Schwarzschild
radius of the black hole with $M$ being the black hole mass) by the
radiation coming from the innermost region of the ADAF will cause an
oscillation of the black hole activity when $L_{\rm acc}\ga
2\%L_{\rm Edd}$. We calculate the durations of the active (or
outburst) and inactive (or quiescent) phases and find that they are
consistent with the observations. The last section (\S4) is devoted
to discussions.

\section{Criticisms on the radiation pressure instability model}

Czerny et al. (2009) assume that the accretion flow in these radio
sources is described by the standard thin disk. We usually believe
that this type of disk suffers from the thermal instability when the
accretion rate is above a threshold, which corresponds to $L\sim
0.025L_{\rm Edd}$  for supermassive black holes (it is $\sim
0.2L_{\rm Edd}$ for stellar-mass black holes)(e.g., Svensson \&
Zdziarski 1994; Janiuk, Czerny \& Siemiginowska 2002). { This
threshold is close to that of the global Compton scattering
instability by coincidence.} Assuming the SEDs of these radio
sources are similar to those of luminous AGNs (refer to \S1), they
estimated their luminosities and found that all the sources in their
sample are above this threshold so they concluded that they are
thermally unstable. As a result of the instability, their accretion
rates undergo high and low values periodically. They further assume
this results in the high and low energy output of the jets. By
adjusting parameters, this model can explain the timescales of both
the active and inactive phases quite well.

However, there are several questions about this model. The first one
is whether jets can be formed from standard thin disks. To answer
this question, let's look at the observations of black hole X-ray
binaries (BHXBs). It is believed that the physics of BHXBs and AGNs
is the same, while the observational data in BHXBs is much better
than the case of AGNs in terms of formation of jets from different
accretion modes. An individual BHXB usually has five states, namely
quiescent, low/hard, intermediate, high/soft, and very high (also
called steep power-law) states. They are characterized by different
spectra and timing features (see a review by McCintock \& Remmilard
2006). Roughly speaking, the features of the soft state are similar
to those of the luminous AGNs while the hard state is similar to
LLAGNs. Different states are believed to be caused by different
accretion modes. Specifically, the accretion flow in the soft state
is believed to be described by the standard thin disk; while that in
the hard state by the hot accretion flow such as an
advection-dominated accretion flow (ADAF; Narayan \& Yi 1994, 1995;
for reviews on ADAFs see Narayan, Mahadevan \& Quataert 1998 and
Narayan \& McClintock 2008) or luminous hot accretion flow (LHAF;
Yuan 2001; Yuan \& Zdziarski 2004). For reviews on the models of
BHXBs and LLAGNs, readers are referred to Zdziarski \& Gierlinski
(2004), Done, Gierlinski \& Kubota (2007), Narayan (2005), and Yuan
(2007). Back to our question of jet formation, radio observations to
BHXBs clearly show that jets exists in the hard state, evidenced by
the flat radio spectrum and the elongated radio image; while both
features disappear in the soft state (Fender 2006). This strongly
indicates that jets can't be formed in the standard thin disk, but
can in the hot accretion flows. Theoretical studies are consistent
with this observational result (e.g., Livio, Ogilvie \& Pringle
1999; Meier 2001).

Observations find that the highest luminosity that hard states can
reach is $\sim (10-30)\%L_{\rm Edd}$. This is roughly consistent
with predictions of hot accretion flows (Yuan \& Zdziarski 2004;
Yuan et al. 2007). { Above this luminosity, the cooling in the hot
accretion flow becomes so strong that the accretion flow can't
remain hot and it collapses.} On the other hand, strong radio
emission has also been detected when BHXBs are more luminous, i.e,
in the very high state. GRS 1915+105 and radio loud quasars may
correspond to this ``radio loud very high state''. But this does not
imply that jets can also be formed in standard thin disks because
the very high state can't be described by a standard thin disk.
Actually, the accretion disk model of the very high state is still
an unsolved problem. More importantly, detailed observations
indicate that the strong radio emission comes from radio-emitting
blobs ejected during the transition from a ``hard very high state''
to a ``soft very high state'', which may be modeled by a hot
accretion flow-like and thin disk-like models, respectively
(McClintock \& Remillard 2006). Such kind of ejection is called
``episodic jets'', to discriminate their many different features
compared to the ``continuous jets'' (see reviews in Fender \&
Belloni 2004; Fender, Belloni \& Gallo 2004). Yuan et al. (2009)
propose a magnetohydrodynamical model for the formation of episodic
jets, by analogy with the coronal mass ejection (CME) phenomenon in
the Sun. Different from the ``continuous jets'' which are formed in
the presence of large-scale open magnetic fields, episodic jets are
formed in the region of closed magnetic fields in the disk corona.
According to this model, episodic jets can be formed in hot
accretion flows and during their collapse into a cold thin disk. So
both direct observations and the theoretical model indicate that
episodic jets are not formed in standard thin disks.

A question is then whether the compact radio sources correspond to
the hard state or very high state. The model proposed in Reynolds \&
Begelman (1997) requires that the duration of the jet launching in
the active phase must be $\sim 10^4$ yr. If the compact radio
sources correspond to the very high state, this duration should be
the timescale of the state transition. Unfortunately, the details of
the state transition are still an open question, although it is
believed to be associated with the collapse of the hot accretion
flow to a cold disk. The observed timescale of the state transition
ranges from a few days to $\sim$ 100 days (Yu \& Yan 2009; Zdziarski
\& Gierlinski 2004). Scaling this timescale with black hole mass by
a factor of $10^7$ gives $10^5-10^6$ yr, which is one or two orders
of magnitude longer than that required by Reynolds \& Begelman
(1997). Another point against the analogy between the compact radio
sources and the very high state is the luminosity. As we state in
\S1, the typical luminosity of these radio sources is $\sim
0.02L_{\rm Edd}$. But on the other hand, the typical transition
luminosity from hard to soft is significantly higher, among the
range of $\sim (2-20)\%L_{\rm Edd}$. So it seems more likely that
these compact radio sources correspond to the hard state rather than
the very high state. But we would like to point out that it is hard
at present to completely rule out the latter possibility.

The second question is on the thermal instability of radiation
dominated thin disk. {As we mentioned at the beginning of this
section, when the accretion rate is higher than a threshold so that
the pressure in the accretion flow is dominated by the radiation
pressure, the standard thin disk was thought to be thermally
unstable (Shakura \& Sunyaev 1976; Piran 1978)}. Time-dependent
{global} hydrodynamical calculations (e.g., Honma et al. 1992;
Janiuk et al. 2002; Li et al. 2007) found that the local thermal
instability will result in the ``limit-cycle'' behavior. However,
Gierlinsk \& Done (2004) show that observations of the soft state of
BHXBs, which has widely been believed to be described by the
standard thin disk model, challenge the above prediction. The
luminosity of the high state sources compiled in their sample spanes
the range $0.01\la L/L_{\rm Edd}\la 0.5$, with the highest one well
exceeding the theoretical stable limit. Thus we should expect to see
the limit-cycle behavior. In contrast to this expectation, however,
observations show little variability, which convincingly indicates
that they are thermally stable. Recently, the non-existence of the
thermal instability of radiation dominated thin disk has also been
directly shown by three-dimensional MHD numerical simulation of
shearing box (Hirose, Krolik \& Blaes 2009).

Some theoretical efforts have been made to solve this puzzle. One
explanation was proposed by Hirose, Krolik \& Blaes (2009). They
argued that although the stress is proportional to the total
pressure\footnote{In this regard, we note that Czerny et al. (2009)
assume that the stress is proportional to the geometrical mean
between the gas and the total pressure.}, the stress fluctuation
precedes pressure fluctuations, contrary to the usual supposition
that pressure controls the magnetic stress. This explains the
thermal stability. An alternative explanation was proposed recently
by Zheng et al. (2011). They argue that the previous analytical
analysis of thermal stability neglect the role of magnetic pressure,
which usually accounts for $\sim 10\%$ of the total pressure thus be
dynamically important. They show that if the magnetic pressure
decreases with response to the increase of temperature of the
accretion flow, the threshold of accretion rate above which the disk
becomes unstable increases significantly compared to the case of not
considering the magnetic pressure. The physical reason is that in
this case the dependence of turbulent dissipation heating on
temperature becomes weaker.

\section{Compton scattering feedback model}

We propose that the intermittent activity of compact radio sources
is caused by the ``global Compton scattering'' feedback effect in
hot accretion flows. The idea of global Compton scattering was first
proposed by Ostriker et al. (1976) and Cowie, Ostriker \& Stark
(1978) in the context of spherical accretion, and later investigated
by Esin (1997), Park \& Ostriker (2001, 2007), and Yuan, Xie \&
Ostriker (2009; hereafter YXO09) for rotating accretion flows. Here
we briefly review the main results of YXO09. Most of the radiation
comes from the innermost region of the accretion flow, $\sim 10r_s$.
The radiation will propagate outward and heat the electrons at large
radii via Compton scattering. If the luminosity is high enough, the
electrons will be heated above their virial temperature thus we
can't obtain the steady solution. In this case, YXO09 argue that the
black hole activity will oscillate between an active and inactive
phases. The quantitative results are as follows. Given the existence
of outflow in hot accretion flows, the mass accretion rate is {a
function of radius, $\dot{M}(r)=\dot{M}_{\rm out} (r/r_{\rm
out})^{s}$, with $\dot{M}_{\rm out}$ being the accretion rate at the
outer boundary $r_{\rm out}$. In YXO09 we adopt $s=0.3$, following
the modeling result of the supermassive black hole in Sgr A* (Yuan,
Quataert \& Narayan 2003). We set $r_{\rm out}=10^5r_s$, since this
is the Bondi radius for typical parameters of black hole mass
$M=10^8\msun$ and interstellar medium temperature $T=10^7 $K, as in
the present work. We find that when $\dot{M}_{\rm out} \ga
\dot{M}_{\rm Edd}$, or equivalently $L\ga 2\% L_{\rm Edd}$, the
electron temperature at $10^5r_s$ will be heated to be higher than
the virial value thus no steady accretion solution exists. This
radius is called virial radius. Its value is found to be roughly
anti-correlated with $\dot{M}_{\rm out}$, $r_{\rm vir}\sim 10^5 r_s
(\dot{M}_{\rm out}/\dot{M}_{\rm Edd})^{-1}$, mainly in the sense
that a larger $\dot{M}_{\rm out}$ corresponds to a smaller $r_{\rm
vir}$. Note that $\dot{M}_{\rm out}$ can't be too small, since in
that case the virial radius will be out of the Bondi radius (i.e.,
the boundary of the accretion flow) thus the calculations in YXO09
may fail}.

In the present work we are satisfied with simple analytical
calculations and estimations, but leave the detailed numerical
simulation to the next step. First we note that the typical
luminosity of the sources compiled in Czerny et al. (2009) is $\sim
2\% L_{\rm Edd}$, consistent with the predicted threshold of
luminosity in YXO09 above which the black hole activity oscillates
due to the global Compton heating. In the following we analyze this
oscillation behavior in more detail, calculating the timescales of
the active and inactive phases and comparing with observations. The
readers are referred to Fig. 1 for the schematic configuration of
the two phases.

We assume the mass of the black hole $M=10^8 M_8\msun$. The density
and temperature of the ISM at Bondi radius are denoted as $n_{\rm
ISM}$ and $T_{\rm ISM}\sim 10^7K$. This high temperature of $T_{\rm
ISM}$ is supported by the direct observations of some galactic
centers, such as Sgr A* (Baganoff et al. 2003), some nearby galaxies
(Pellegrini 2005), and the radio galaxy M~87 (e.g., Di Matteo et al.
2003). Theoretically, this could be because of the shock heating due
to the collision between stellar winds (Quataert 2004). The outer
boundary of the accretion flow is set at the Bondi radius, $r_{\rm
out}\sim r_B\sim GM/c_s^2\sim 10^5 r_s$.

As we state above, when $L\sim 2\%L_{\rm Edd}$ (or equivalently
$\dot{M}_{\rm out}\sim \dot{M}_{\rm Edd}$), the gas at $r_{\rm
vir}\sim r_{B}$ will be heated to be above the virial temperature
thus will diffuse. The gas within $r_{\rm vir}$ can still be
accreted, which powers the high-luminosity active phase. The
duration of this phase is determined by the accretion timescale at
$r_{\rm vir}$, \be t_{\rm act}\sim t_{\rm acc}(r_{\rm vir})\sim
\frac{r_{\rm vir}}{v_r} \sim 3\times
10^4M_8\left(\frac{\alpha}{0.1}\right)^{-1} \left(\frac{\dot{M}_{\rm
out}}{\dot{M}_{\rm Edd}}\right)^{-1} {\rm yr}\ee (Narayan, Mahadevan
\& Quataert 1998). For typical values of $\dot{M}_{\rm out}\approx
\dot{M}_{\rm Edd}$, as indicated by the observed $L\sim 2\%L_{\rm
Edd}$, this timescale is in good agreement with the required
duration of active phase in Reynolds \& Begelman (1997). Moreover,
the sample compiled in Czerny et al. (2009) indicates that sources
with higher luminosity have younger age. This is easy to understand
from eq. (1), because for an ADAF we roughly have $L\propto
\dot{M}_{\rm out}^2$ thus $t_{\rm act}\sim 3\times 10^4
M_8(0.1/\alpha)(L/0.02L_{\rm Edd})^{-1/2}~{\rm yr}$.

The temperature profile within $r_{\rm vir}\sim 10^5r_s$ has been
calculated in YXO09. Now we estimate the temperature of the gas
beyond $10^5r_s$, which is required to calculate the duration of the
inactive phase. The temperature is mainly determined by the Compton
heating, viscous dissipation, and bremsstrahlung cooling (we don't
consider other possible heating mechanisms for simplicity). Cooling
is usually negligible in an ADAF. In the non-relativistic limit, the
Compton heating reads (YXO09), \be q_{\rm comp}=n_e
(\theta_x-\theta_e)F\sigma_T\approx n_e\theta_xF\sigma_T\propto
r^{-3.2}. \ee Here $\theta_x\equiv k T_x/m_ec^2$ and $T_x$ is the
radiation temperature characterizing the spectrum of radiation from
the ADAF, $F$ is the radiation flux, $\sigma_T$ is Thomson cross
section. The value of $T_x$ can be well approximated to be (YXO09),
\be T_x\sim 10^9 ~{\rm K}.\ee Note that this value is much larger
than the typical value for the spectrum of luminous quasar, where
$T_x\sim 10^7{\rm K}$ (Sazonov et al. 2005). This is of course
because of the significant difference between the SEDs of luminous
AGNs and LLAGNs (Ho 1999; Ho 2008; Yuan, Xie \& Ostriker 2009). The
viscous heating rate is given by \be q_{\rm vis}\propto n_e
\theta_i^{1/2}Hr^2\left(d\Omega/dr\right)^2\propto r^{-3.7}.\ee In
both eqs. (2) and (4), the scaling of an ADAF (with outflow)
$n_e\propto r^{-3/2+s}=r^{-1.2}$ and $\theta_i\equiv k
T_i/m_pc^2\propto r^{-1}$ are adopted. At $r\sim 10^5r_s$, we have
$q_{\rm comp}\ga q_{\rm vis}$, since this is why the electrons are
heated to be above the virial temperature. Eqs. (2) and (4) then
imply that the heating is dominated by Compton scattering at $r\ga
10^5 r_s$. The evolution of electron temperature is then described
by \be n_ek\frac{dT_e}{dt}=n_e(\theta_x-\theta_e)F\sigma_T. \ee The
solution is \be T_e=T_{e0}e^{-At}+T_x(1-e^{-At}).\ee Here $T_{e0}$
is the electron temperature at the beginning of the active phase and
$A=\frac{L\sigma_T}{4\pi r^2 m_e c^2}$. The heating timescale is the
duration of the active phase $t_{\rm act}$, so the spatial range
that electrons can be heated to $T_x$ is determined by \be At_{\rm
act}\approx 1.\ee This gives \be r\sim \left(\frac{L\sigma_Tt_{\rm
act}}{4\pi m_ec^2}\right)^{1/2}\sim 10^6 r_s\ee In the last
calculation the typical values of $L\sim 2\%L_{\rm Edd}$ and $t_{\rm
act}$ from eq. (1) are adopted. So we obtain that the gas up to
$\sim 10^6r_s$ can be heated to Compton temperature $T_x$ after the
active phase. Note that we neglect the bremsstrahlung cooling in our
calculation, so this value should be regarded as an upper limit.

In correspondence with the increase of the temperature, the density
of the gas should become much lower than the unheated ISM. The
accretion of these low-density gas results in the significant
decrease of mass accretion rate and the luminosity. The source then
enters into the inactive phase. We denote the density as $n_{\rm
inact}$ and estimate its value by the pressure balance of these gas
with the surrounding ISM, whose density and temperature are denoted
as $n_{\rm ISM}$ and $T_{\rm ISM}(\sim 10^7{\rm K})$: \be n_{\rm
inact}=n_{\rm ISM}\frac{T_{\rm ISM}}{T_x}=10^{-2}n_{\rm ISM}. \ee We
estimate the mass accretion rate with the Bondi accretion theory, so
\be r_{\rm Bondi}\sim \frac{GM}{c_s^2},\ee and \be \dot{M}_{\rm
inact}\sim \alpha\dot{M}_{\rm Bondi}\approx 4\pi \alpha r_{\rm
Bondi}^2n_{\rm inact} m_p c_s(r_{\rm Bondi})\propto T^{-3/2}n_{\rm
inact}.\ee Here $\alpha$ is the viscous parameter, $c_s(r_{\rm
Bondi})$ is the sound speed at $r_{\rm Bondi}$. So the accretion
rate is up to $\sim 10^5$ times lower in the inactive phase compared
to the active phase, which is why the source becomes inactive.

The Compton heated gas from $r_{\rm Bondi}$ to $10^6 r_s$ will cool
with time. The duration of the inactive phase should be determined
by the depletion of the heated gas due to accretion, or the cooling
of these heated gas, whichever is shorter. The former is hard to
estimate since $10^6 r_s$ is well out of the influence sphere of the
black hole. But it must be much longer than the accretion timescale
determined by the ADAF theory, i.e., \be t_{\rm acc} \gg \frac{10^6
r_s}{v_r}\sim 10^5 M_8\left(\frac{\alpha}{0.1}\right)^{-1}~{\rm
yr}.\ee The cooling timescale reads
$$ t_{\rm cool}=\frac{n_{\rm inact} kT_x}{j_{\rm brem}}=6\times
10^{3} T_x^{1/2}n_{\rm inact}^{-1} ~{\rm yr}.$$ Here $j_{\rm brem}$
is the bremsstrahlung emissivity. We have $T_x=10^9{\rm K}$ and
$n_{\rm inact}=10^{-2}n_{\rm ISM}$, but what is the value of $n_{\rm
ISM}$? From our statement above, we know that the condition for the
oscillation of the black hole activity is that the luminosity of the
active phase is $L\ga 2\%L_{\rm Edd}$, or equivalently
$\dot{M}(10^5r_s)\ga\dot{M}_{\rm Edd}$. Given that $\dot{M}_{\rm
out}(10^5r_s)\sim\alpha\dot{M}_{\rm Bondi}=4\pi \alpha r_{\rm
Bondi}^2c_s n_{\rm ISM}m_p$, we obtain $n_{\rm ISM}\ga
10^5(\alpha/0.1)^{-1}M_8^{-1}{\rm cm}^{-3}$. So we have \be t_{\rm
cool}=10^5 M_8
\left(\frac{\alpha}{0.1}\right)\left(\frac{T_x}{10^9{\rm
K}}\right)^{1/2}~{\rm yr}.\ee This determines the duration of the
inactive phase because it is much shorter than the accretion
timescale at $10^6r_s$ (eq. 12). This value is in good agreement
with the duration of the inactive phase required in Reynolds \&
Begelman (1997).

\section{Discussion: comparisons with other works}

In this paper we propose an AGN feedback mechanism occurring at
$r_{\rm vir}\sim 10^5r_s$ from the central black hole, i.e., around
parsec scale. The conditions for this effect to be present are that
the accretion flow is described by hot accretion flow (such as an
ADAF), and the accretion luminosity $L\ga 2\%L_{\rm Edd}$. {\em The
influence of this effect is that the black hole will oscillate
between an active and inactive phases, which last for $3\times 10^4
(0.1/\alpha)(M/10^8\msun) (L/2\%L_{\rm Edd})^{-1/2}~{\rm yr}$ and
$10^5(\alpha/0.1)(M/10^8\msun)~{\rm yr}$, respectively.} Such kind
of short-timescale oscillation should be common in radio sources
with $L\ga 2\%L_{\rm Edd}$. This result implies that even though the
fueling rate of AGNs is constant at large scales, the radiation (and
also matter) output from the AGNs oscillates. This behavior should
be taken into account when we consider the effects of AGNs feedback
at larger scales, especially those treating the small-scale
accretion as the ``sub-grid'' input of their simulations.

Now let's compare our study with other works. One is the series of
works by Ciotti \& Ostriker (2001; 2007). In their works, they focus
on the radiative heating by the central AGN to the ISM, but on
scales much larger than ours, from several pc to kpc. More physics,
such as supernova heating, are included there. Similar to our
results, they also find that when the luminosity of the central AGNs
is high, in their case close to $L_{\rm Edd}$, oscillation occurs.
However, the timescale of oscillation is much longer, $\sim
(10^7-10^8) ~{\rm yr}$. One important reason is that the radiation
temperature adopted in their model is much smaller, $T_x\sim 10^7
{\rm K}$. Physically, such a low $T_x$ corresponds to the typical
spectrum of a quasar (Sazanov et al. 2005). While when the accretion
rate is lower than a certain value, the accretion flow should make a
transition from the standard thin disk to an ADAF, as argued in
Narayan \& Yi (1995) and evidenced by the transition from a soft to
a hard states in BHXBs. The spectrum emitted by ADAFs is
significantly different from that of quasars, most significantly
characterized by the lack of big-blue-bump, as we point out in \S1
(Ho 1999; Ho 2008). The corresponding radiation temperature of such
spectra is typically $T_x\sim 10^9{\rm K}$ (YXO09).  Another
possibly important point is the inner boundary of the simulation
domain. From YXO09, the global Compton scattering effect is already
important at $\sim 10^5r_s(\dot{M}/\dot{M}_{\rm Edd})^{-1}$, i.e, at
sub-pc or pc scales. But in Ciotti \& Ostriker's works, they
typically set the inner boundary at several pc. Thus some important
feedback effect may be missed. Therefore, it will be interesting to
repeat the simulations of Ciotti \& Ostriker (2001; 2007), but by
taking into account the change of $T_x$ at low $\dot{M}$ and the
location of the inner boundary. This work is in progress.

Another series of works are by Proga and his collaborators (e.g.,
Proga, Stone \& Kallman 2000; Proga 2007; Kurosawa \& Proga 2009).
Similar to our work, they focus on the sub-pc and pc scales.
Different from our work, but similar to Ciotti \& Ostriker (2001;
2007), they assume that the central radiation spectrum is
quasar-like. Another difference from our work is that they focus on
cold accretion flows and higher mass accretion rates. However, since
oscillation behavior was also found in their works, if the accretion
rate decreases to such a low value that the accretion flow makes a
transition from a thin disk to a hot accretion flow, $T_x\sim 10^9
{\rm K}$ should again be adopted.


\section*{Acknowledgments}

We thank Bozena Czerny and the anonymous referee for helpful
comments. This work was supported in part by the Natural Science
Foundation of China (grants 10833002, 10821302, and 10825314), the
National Basic Research Program of China (973 Program 2009CB824800),
and the CAS/SAFEA International Partnership Program for Creative
Research Teams.

\clearpage

\begin{figure}
\epsscale{0.3}\plotone{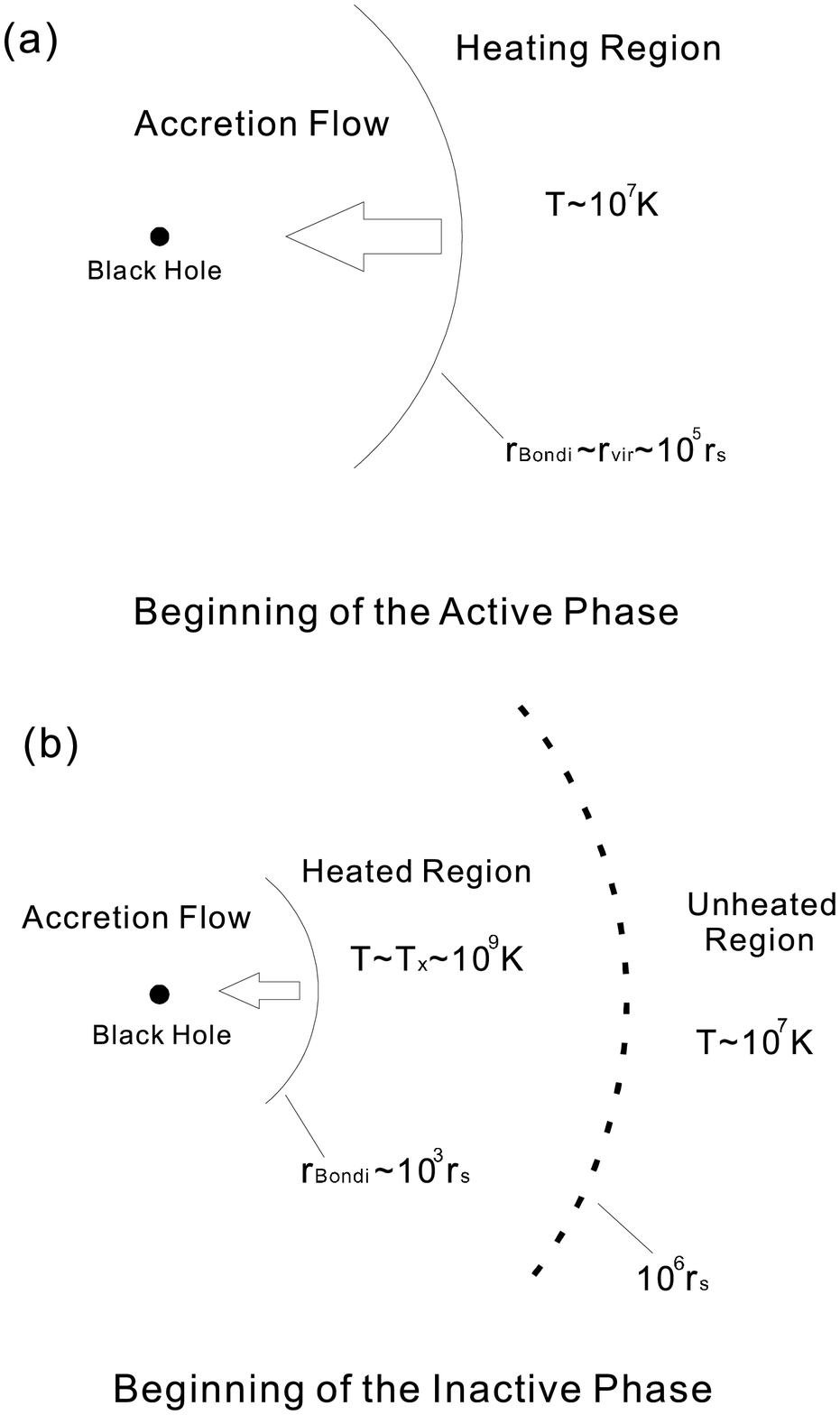}\hspace{15cm}\vspace{1cm} \caption{The
schematic figure of the beginning of the active (a) and inactive (b)
phases. At the active phase, the accreting gas inside $10^5r_s$ is
virial; the gas outside this radius is heated by the global Compton
scattering toward $T_x\sim 10^9{\rm K}$, which is higher than the
virial value. Thus this phase finishes when the gas inside $10^5r_s$
is depleted; the source then enters into the inactive phase. At the
inactive phase, the gas is very hot, with $T\sim T_x\sim 10^9{\rm
K}$, thus the mass accretion rate and the luminosity is very low.
This phase finishes when the hot gas cools down. }
\end{figure}


\begin{thebibliography}{}

\bibitem[]{} Baganoff, F.~K. 2003, ApJ, 591, 891


\bibitem{} Ciotti, L. \& Ostriker, J.P. 2001, ApJ, 551, 131

\bibitem{} Ciotti, L. \& Ostriker, J.P. 2007, ApJ, 665, 1038

\bibitem[]{} Cowie, L.~L., Ostriker, J.~P., \& Stark, A.~A. 1978, ApJ, 226, 1041

\bibitem[]{} Croton, D.~J. et al. 2006, MNRAS, 365, 11

\bibitem{} Czerny, B. et al. 2009, ApJ, 698, 840

\bibitem[]{} Di Matteo, T. et al. 2003, ApJ, 582, 133

\bibitem[]{} Di Matteo, T., Springel, V., \& Hernquist, L. 2005,
Nature, 433, 604

\bibitem{} Done, C., Gierli\'nski, M., \& Kubota, A. 2007, A\&ARev., 15, 1


\bibitem[]{} Elvis, M. et al. 1994, ApJS, 95, 1

\bibitem[]{} Esin, A. 1997, ApJ, 482, 400

\bibitem[]{} Fabian, A.C. 1999, MNRAS, 308, L39

\bibitem[]{} Fender R.~P., In: Compact stellar X-ray sources.
Cambridge Astrophysics Series, 2006, p. 381 (astro-ph/0303339)

\bibitem[]{} Fender R.~P., Belloni T.~M., 2004, \araa, 42, 317

\bibitem{FBG04}  Fender R.~P., Belloni T.~M., Gallo E., 2004, \mnras, 355, 1105

\bibitem{} Gierli\'nski, M., \& Done, C. 2004, \mnras, 347, 885

\bibitem[]{} Giroletti, M., Giovannini, G., \& Taylor, G.B. 2005,
A\&A, 474, 409

\bibitem{} Hirose, S., Krolik, J. H., \& Blaes, O. 2009, \apj, 691, 16

\bibitem[]{} Ho, L. 1999, \apj, 516, 672

\bibitem[]{} Ho, L. 2008, ARA\&A, 46, 475

\bibitem[]{} Ho, L. 2009, \apj, 699, 626

\bibitem[]{} Honma, F., Matsumoto, R., \& Kato, S. 1992, \pasj, 44, 529

\bibitem[]{} Janiuk, A., \& Czerny, B. 2011, arXiv:1102.3257

\bibitem[]{} Janiuk, A., Czerny, B., \& Siemiginowska, A. 2002, \apj, 576, 908

\bibitem[]{} Kaiser, C. R., Schoenmakers, A. P., \& Roettgering, H. J. A. 2000,
MNRAS, 315, 381

\bibitem[]{} King, A. R. 2003, ApJ, 596, L27

\bibitem[]{} Kunert-Bajraszewska, M., Marecki, A., Thomasson, P., \& Spencer,
R.~E. 2005, A\&A, 440, 93

\bibitem[]{} Kurosawa, R. \& Proga, D. 2009, MNRAS, 397, 1791

\bibitem[]{} Li, S. -L., Xue, L., \& Lu, J. -F. 2007, \apj, 666, 368

\bibitem[]{} Livio, M., Ogilvie, G.~I., \& Pringle, J.~E, 1999, ApJ,
512, 100

\bibitem[]{} McClintock, J. E., \& Remillard, R. A. 2006, in Compact
  Stellar X-Ray Sources, ed. W. H. G. Lewin \& M. van der Klis,
  (Cambridge: Cambridge Univ.\ Press), 157

\bibitem[]{} Meier, D.~L. 2001, ApJ, 548, L9

\bibitem[]{} Murray, N., Quataert, E., \& Thompson, T.A. 2005, ApJ,
618, 569

\bibitem[]{} Narayan, R. 2005, Ap\&SS, 300, 1

\bibitem{n98} Narayan R., Mahadevan R. Quataert E., 1998, in
Theory of Black Hole Accretion Disks, eds.\ M.~A. Abramowicz, G.
Bjornsson, and J.~E. Pringle. Cambridge University Press, p.\ 148

\bibitem[]{} Narayan, R., \& McClintock, J.E. 2008, NewAR, 51, 733

\bibitem[]{} Narayan, R., \& Yi, I. 1994, ApJ, 428, L13

\bibitem[]{} Narayan, R., \& Yi, I. 1995, ApJ, 452, 710

\bibitem[]{} O\'Dea, C. P., \& Baum, S. A. 1997, AJ, 113, 148

\bibitem[]{} Ostriker, J.~P., McCray, R., Weaver, R. \& Yahil, A.
1976, ApJ, 208, L61

\bibitem[]{} Ostriker, J.P. et al. 2010, ApJ, 722, 642

\bibitem[]{} Park, M. \& Ostriker, J.~P. 2001, ApJ, 549, 100

\bibitem[]{} Park, M. \& Ostriker, J.~P. 2007, ApJ, 655, 88

\bibitem[]{} Parma, P. et al. 2007, A\&A, 470, 875

\bibitem[]{} Pelligrini, S. 2005, ApJ, 624, 155

\bibitem[]{} Piran, T. 1978, \apj, 221, 652

\bibitem[]{} Proga, D. 2007, ApJ,  661, 693

\bibitem[]{} Proga, D., Stone, J.~M., \& Kallman, T.~R. 2000, ApJ,
543, 686

\bibitem[]{} Quataert, E. 2004, ApJ, 613, 322

\bibitem[]{} Reynolds, C. S., \& Begelman, M. C. 1997, ApJ, 487, L135

\bibitem[]{} Sazonov, S.~Yu., Ostriker, J.~P., Ciotti, L., \& Sunyaev,
R.~A. 2005, MNRAS, 358, 168

\bibitem[]{} Siemiginowska, A. et al. 2008, ApJ, 684, 811

\bibitem[]{} Silk, J., \& Rees, M. 1998, A\&A, 331, L1

\bibitem[]{} Snellen, I.~A.~G., et al. 2000, MNRAS, 319, 445

\bibitem[]{} Starwarz, L. et al. 2004, ApJ, 613, 119

\bibitem[]{} Svensson, R. \& Zdziarski, A.A. 1994, ApJ, 436, 599

\bibitem[]{} Vink, J. et al. 2006, MNRAS, 367, 928

\bibitem[]{} Yu, W. \& Yan, Z. 2009, ApJ, 701, 1940

\bibitem[]{} Yuan, F. 2001, MNRAS, 324, 119

\bibitem[]{} Yuan, F. 2007, in The Central Engine of Active
Galactic Nuclei, ed. L. C. Ho \& J.-M. Wang (San Francisco: ASP), 95

\bibitem[]{} Yuan, F., Lin, J., Wu, K. \& Ho, L. 2009, MNRAS, 395,
2183

\bibitem[]{} Yuan, F., Xie, F. \& Ostriker, J.~P. 2009, ApJ, 691,
98 (YXO09)

\bibitem[]{} Yuan, F., Zdziarski, A.A., Xue, Y.~Q., \& Wu, X.~B. 2007, ApJ,
659, 541

\bibitem[]{} Zdziarski, A. A., \& Gierli\'nski, M. 2004,  Prog. Theor. Phys.
Suppl., 155, 99

\bibitem[]{} Zheng, S., Yuan, F., Gu, W.~M., \& Lu, J.~F. 2011, ApJ,
732, 52

\end{thebibliography}
\end{document}